\documentclass[preprint]{revtex4-1}

\usepackage[]{graphicx}

\begin{document}

\title{Two-step Dynamics of Photoinduced Phonon Entanglement Generation between Remote Electron-phonon Systems}
\author{Kunio Ishida}
\email{ishd\_kn@cc.utsunomiya-u.ac.jp}
\affiliation{School of Engineering and Center for Optical Research and Education, Utsunomiya University, Utsunomiya, Tochigi 321-8585, Japan}
\author{Hiroaki Matsueda}
\affiliation{Department of Applied Physics, Graduate School of Engineering, Tohoku University, Sendai 980-8579, Japan}

\begin{abstract}
  The generation of quantum entanglement between phonons in photoirradiated remote electron-phonon systems is numerically studied.
  Upon excitation by a visible/ultraviolet laser pulse, the entanglement of electrons is immediately generated and that of phonons follows via electron-phonon interactions, i.e., the entanglement generation of phonons is a two-step process.
  Therefore, it is important to design the temporal properties of incident optical pulses in order to control the entanglement of electrons and/or phonons.
  These features are revealed by the quantum mutual information and the composite modes derived from the Heisenberg equation of motion.
  The calculated results also show that the dynamics of the phonon entanglement can be observed by time-resolved spectroscopy on the scattered light.
\end{abstract}

\maketitle

\section{Introduction}
\label{intro}
As the recent progress of ultrashort pulse laser technology has made it possible to observe the time evolution of quantum-mechanical states in the coherent regime, the transient properties of condensed matter have been studied by various time-resolved techniques\cite{measure2,measure3,measure4,measure5}.
When these techniques are considered from application side, light control of quantum many-body states has become an intriguing problem, which motivated us to pay attention to the quantum nature of light-matter interaction as well.

Considering that the injection of optical excitation induces cooperative phenomena of electrons and/or lattices\cite{pipt1,pipt2,pipt3}, the photoinduced quantum phase transition is a next step in the coherent control of condensed matter.
Furthermore, we should point out that quantum entanglement is a basic notion in the quantum information technology, which also gives an insight into various aspects of quantum many-body states\cite{entangle1,entangle2}.
Therefore, it is important to reveal the mechanism of cooperative phenomena on excited-state manifolds by means of the quantum information theory in order to manipulate the many-body states in coherent regime.
These types of quantum-mechanical control are reminiscent of ``quantum materials'' realized as transient states of materials under an external field\cite{qm}, where profound knowledge of quantum many-body states is required to obtain a material design method in the transient regime.
Summarizing these points, we point out that we require a theory of coherent dynamics of entanglement generation referring to the previous studies on the entanglement in stationary states\cite{ciancio,sato}.

On the other hand, the generation and storage of quantum entanglement between remote systems has attracted our attention, where nonlocal correlation between qubits plays an essential role.
Recent experimental studies exemplified methods of light-mediated entanglement generation between non-interacting systems, which showed
that ultrashort laser pulses generate entanglement between phonons (phonon entanglement) in diamond crystals\cite{lee}, and that quantum correlation between distant quantum dots is induced by microwave irradiation\cite{borjans}.
Although the manipulation of multi-qubit systems has only been performed at low temperatures\cite{borjans},  phonon entanglement can be manipulated at room temperature\cite{lee}, which is advantageous for practical applications.
They also provide us  a clue for understanding the relationship between the quantum entanglement and the dynamical control of many-body states in materials.
Furthermore, since many of photoinduced cooperative phenomena is concomitant with the structural change of molecules/crystals\cite{piptme}, electron-phonon interaction plays an important role in the coherent dynamics of  quantum phase transition under light irradiation.

In this study, we numerically investigate the dynamics of photoinduced quantum entanglement generation in remote electron-phonon systems.
We focus on the dynamics of phonon entanglement generation, which is also a clue to understand the initial processes of photoinduced structural change, e.g., formation of embryonic nuclei\cite{piptme2}.
Furthermore, as shown in previous studies on the Tavis-Cummings model\cite{tc} of multi-qubit entanglement\cite{tc1,tc2,tc3} and spin-chain dynamics\cite{schain}, quantized light is necessary to describe photoinduced entanglement generation, and thus we should consider a model of electron-phonon-photon systems.

Since the entire system is composed of multiple types of quanta, it is not appropriate to compare the dynamical behavior of the entire system with that of qubit systems.
To be more precise, when phonons are created by Raman scattering, the quantum correlation between those in different systems is mediated by excited electrons via electron-phonon interaction.
We found that phonon entanglement follows the electron entanglement with a certain delay corresponding to the creation time of phonons.
We discuss the mechanism of entanglement generation by using the Heisenberg equations of motion.

We also study the measurement of entanglement properties which is also of interest from an experimental viewpoint.
Since scattered light carries information on the nonlocal correlation between material systems,  we discuss the detection of entanglement by time-resolved spectroscopy.
We found that the heterodyne detection of the Stokes light makes it possible to access the two-step dynamics of phonon entanglement generation.

\section{Model and Method}

\subsection{Hamiltonian of the System}
\begin{figure}
 \scalebox{1}{\includegraphics*{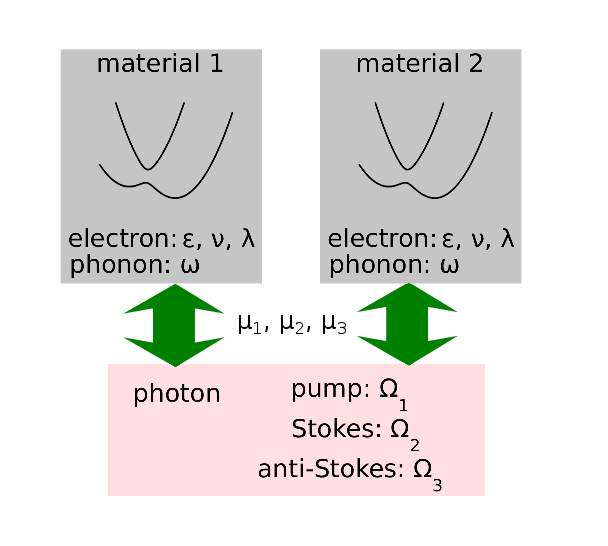}}
\caption{(Color online) A schematic view of the model.}
\label{mdfig}
\end{figure}

We aimed to calculate the quantum dynamics of multiple electron-phonon systems irradiated by photons.
For this purpose, we employ a model of $M$ non-interacting electron-phonon systems.
When all of the systems interact with multimode photons, the Hamiltonian is described by\cite{epjdme}
\begin{eqnarray}
  {\cal H} & = & \sum_{i=1}^3 \Omega_i c_i^\dagger c_i \nonumber \\
           & + & \sum_{j=1}^M \left [ \omega  a_j^\dagger a_j + \sigma_x^j \left \{ \sum_{i=1}^3 \mu_i (c_i^\dagger + c_i ) + \lambda \right \} \right . \nonumber \\
& + & \left . \hat{n}_j \{ \nu (a_j^\dagger + a_j) + \varepsilon \}\right ],
\label{ham}
\end{eqnarray}
where $a_j$ and $c_i$ denote the annihilation operators of optical phonons in the $j$-th material system (system $j$) and the photons of the $i$-th mode, respectively.
$\sigma_\rho^j\ (\rho = x,y,z)$ describes the Pauli matrix that operates on the electronic states of the $j$-th material system, $|g \rangle_j$(ground state) and $|e \rangle_j$(excited state).
$\hat{n}_j=(\sigma_z^j +1)/2$ is the excited state electronic population of the $j$-th system, and $\mu_i$ denotes the dipole interaction between electrons and photons.
We take into account the nonadiabatic interaction between the electronic states $\lambda \sigma_x^j$, which contributes to the entanglement generation as well as the other parameters.
Figure \ref{mdfig} shows a schematic view of the model.

We take $\hbar =1$ and the parameter values are $\omega=1$, $\mu_i=0.5 (i=1,2,3)$, $\lambda=1.5$, and $\varepsilon=13.5$.
We also set the electron-phonon coupling (Huang-Rhys factor) $\nu$ to 3.5, a typical value for various materials\cite{inorg,org}. 
Since Raman scattering is one of the primary processes in the excitation of electron-phonon systems, three photon modes corresponding to the pump mode ($\Omega_1=13.5$), Stokes mode ($\Omega_2=12.5$), and anti-Stokes mode ($\Omega_3=14.5$) are taken into account.
In this paper, we consider the dynamics of structural change accompanied by nonadiabatic transition, e.g, photoisomerization, though we do not restrict ourselves to discuss molecular systems.
We also stress that the present model covers various types of photoinduced phenomena, e.g., coherent phonon dynamics\cite{measure2}, conical intersection dynamics\cite{CI}, and photoinduced nucleation in coherent regime\cite{piptme}.

The time-dependent Schr\"odinger equation for Hamiltonian (\ref{ham}) was numerically solved for $M=2$ to obtain the wavefunction $|\Phi(t)\rangle$.
$|\Phi(t)\rangle$ is calculated by the fourth order Runge-Kutta method, where the time step is set to $5\times 10^{-5}$.
The dimension of the  photon and phonon Hilbert spaces are truncated to 63 and 31 per mode, respectively, and thus the dimension of the entire Hilbert space is $2^{30}$ for $M=2$.
We confirmed the convergence of the calculated results.

The initial condition is given by $|\Phi(0) \rangle = |\alpha_1\rangle \otimes |\alpha_2\rangle \otimes |\alpha_3\rangle \otimes |G\rangle_1 \otimes|G\rangle_2 $, where $|\alpha_i \rangle$ denotes a coherent state parameterized by $\alpha_i$ for the $i$-th photon mode and $|G \rangle_j$ is the ground state of the $j$-th electron-phonon system.
In the present study we set $\alpha_1=5$ and $\alpha_2=\alpha_3=-2.5$ in order to describe a train of optical pulses.
The expectation value of the electric field of the incident optical pulse is described by $E(t) \propto \sum_{n=1}^3 i\langle \alpha_ne^{-i\omega_nt} |(c_n-c_n^\dagger)|\alpha_ne^{-i\omega_nt}\rangle \propto  \sum_{n=1}^3 \alpha_n \sin \omega_nt$.

Since the phonons in systems 1 and 2 compose a subsystem of the electron-phonon-photon system, we must consider multipartite entanglement to study phonon entanglement.
Hence, we discuss an appropriate measure for phonon entanglement.
Since the concurrence or tangle is only applicable to the evaluation of multi-qubit entanglement\cite{tc1,tc2,tc3}, we instead discuss the quantitative properties of phonon entanglement by two different methods.
First, we calculate the quantum mutual information, which is interpreted as the relative entropy of the reduced density matrix for the phonon subspace to its closest separable state\cite{qmi1,qmi2}, and is also an order parameter in quantum phase transitions\cite{qpt1}.
In the second method we use a set of composite modes derived from the Heisenberg equation of motion\cite{comp1,comp2}, and studied the mechanism of phonon entanglement generation by comparing the results of the two methods.

\subsection{Quantum Mutual Information}

  When the entire system is divided into three subsystems $A$, $B$, and $C$, the quantum mutual information between $A$ and $B$ is defined by
\begin{equation}
  I_M = S_A + S_B - S_{A\otimes B},
\end{equation}
where $S_\alpha$ is the von Neumann entropy of the reduced density matrix on subsystem $\alpha$.
The wavefunction of the entire system is expressed as 
\begin{equation}
  |\Phi \rangle = \sum_{lmn k k'\sigma \sigma'} p_{lmnkk'\sigma \sigma'} |l\rangle|m\rangle|n\rangle|k \sigma \rangle |k' \sigma\rangle,
\end{equation}
where $|l\rangle$,$|m\rangle$, and $|n\rangle$ denote the Fock state for the photons of each mode.
$|k\sigma\rangle$ shows the $k$ phonon state in electronic level $\sigma (=+,-)$.
When we calculate the phonon entropy, $A$ and $B$ corresponds to the phonon states in systems 1 and 2, respectively.
In this case $C$ denotes the residual degrees of freedom including photons, electrons, and the irrelevant phonons.
Hence, the reduced density matrices for calculation of $S_A$, $S_B$, and $S_{A\otimes B}$  are given by
\begin{eqnarray}
  \rho_A & = & {\rm Tr}_{B\otimes C}|\Phi \rangle \langle \Phi | = \sum_{k,k'} q_{kk'}|k\rangle \langle k'|, \\
  \rho_B & = & {\rm Tr}_{A \otimes C}|\Phi \rangle \langle \Phi | = \sum_{k,k'} r_{kk'}|k\rangle \langle k'|, \\
  \rho_{A \otimes B} & = & {\rm Tr}_C|\Phi \rangle \langle \Phi | = \sum_{k,k',j,j'} s_{kk'jj'}|kj\rangle \langle k'j'|,
\end{eqnarray}
where
\begin{eqnarray}
  q_{kk'}= \sum_{lmnj\sigma\sigma'}\bar{p}_{lmnkj\sigma\sigma'}p_{lmnk'j\sigma\sigma'},\\
  r_{kk'}= \sum_{lmnj\sigma\sigma'}\bar{p}_{lmnjk\sigma\sigma'}p_{lmnjk'\sigma\sigma'},\\
  s_{kk'jj'}= \sum_{lmn\sigma\sigma'}\bar{p}_{lmnkj\sigma\sigma'}p_{lmnk'j'\sigma\sigma'}.
\end{eqnarray}
$\bar{x}$ denotes the complex conjugate of $x$ and ${\rm Tr}_{X}$ shows the partial trace over the basis of system $X$.

When we calculate the electron entropy, $A$ and $B$ corresponds to the electronic states in systems 1 and 2, respectively.
Thus, $C$ denotes the residual degrees of freedom including photons, phonons, and the irrelevant electron.
The reduced density matrices for calculation of $S_A$, $S_B$, and $S_{A\otimes B}$  are given by
\begin{eqnarray}
  \rho_A & = & {\rm Tr}_{B\otimes C}|\Phi \rangle \langle \Phi | = \sum_{\sigma,\sigma'} u_{\sigma \sigma'}|\sigma \rangle \langle \sigma'|, \\
  \rho_B & = & {\rm Tr}_{A \otimes C}|\Phi \rangle \langle \Phi | = \sum_{\sigma,\sigma'} v_{\sigma \sigma'}|\sigma \rangle \langle \sigma k'|, \\
  \rho_{A \otimes B} & = & {\rm Tr}_{C}|\Phi \rangle \langle \Phi | = \sum_{\sigma, \sigma',\tau,\tau'} w_{\sigma \sigma' \tau \tau'}|\sigma \tau\rangle \langle \sigma'\tau'|,
\end{eqnarray}
where
\begin{eqnarray}
  u_{\sigma \sigma'}= \sum_{lmnkk'\tau}\bar{p}_{lmnkk'\sigma\tau}p_{lmnkk'\sigma' \tau},\\
  v_{\sigma \sigma'}= \sum_{lmnkk'\tau}\bar{p}_{lmnkk'\tau\sigma}p_{lmnkk'\tau \sigma'},\\
  w_{\sigma \sigma' \tau \tau'}= \sum_{lmnkk'}\bar{p}_{lmnkk'\sigma\tau}p_{lmnkk' \sigma' \tau'}.
\end{eqnarray}

The quantum mutual information is obtained by
\begin{equation}
  I_M = -{\rm Tr}\rho_A \ln \rho_A-{\rm Tr}\rho_B \ln \rho_B+{\rm Tr}\rho_{A\otimes B} \ln \rho_{A \otimes B}.
\end{equation}

\subsection{Heisenberg Equations of Motion for the Scattered Light}

Although the quantum mutual information is an appropriate measure for phonon entanglement, it is not directly linked to any observables.
In order to study the measurement of entanglement dynamics, we calculated the dynamics of the scattered light which carries information on quantum correlations.
We show that the detection of the scattered light enables us to discuss the relationship between $I_M(t)$ and physical properties.

However, since the signal of the phonon quantum correlation on the scattered light is weak, we consider the heterodyne detection of the output light pulse for which the frequency of the reference light is $\Omega_1$.
Hence, we calculated the signal on the Stokes light, which is proportional to $\sqrt{\Omega_2}{\rm Re}(\langle c_2 \rangle e^{i\Omega_1 t})$.
$\sigma_x^j(t)$, however, also contains the dynamical properties of individual electron-phonon systems that are irrelevant to the entanglement between the two systems.
Therefore, we subtract the corresponding signal for $M=1$ from that for $M=2$.

Moving on to the Heisenberg picture, $c_2(t) = \exp(i{\cal H}/t)c_2\exp(-i{\cal H}/t)$ satisfies the equation of motion given by
\begin{equation}
  i{dc_2(t) \over dt} = [c_2(t),{\cal H}] = \Omega_2 c_2(t) + \mu_2 \sum_{j=1}^2 \sigma_x^j(t),
  \label{eqm}
\end{equation}
where the operators with explicit time dependence are expressed in the Heisenberg representation.
The first term of the rhs of Eq.\ (\ref{eqm}) describes the beat of the photons, while the second term represents the effect of the electron-photon (dipole) interaction.
Hence, the quantum correlation between systems 1 and 2 is reflected in the Stokes light through $\sigma_x^j(t)$.

  The formal solution of Eq.\ (\ref{eqm}) is given by
  \begin{equation}
    c_i(t) = c_i e^{-i\Omega_i t} - i\mu_i e^{-i\Omega_i t} \sum_j \int_0^t \sigma_x^j(s) e^{i\Omega_i s} ds.
  \end{equation}
  
Since the complex amplitude of the scattered light $E(t)$ is proportional to $i\sqrt{\Omega_i }  \langle c_i(t) \rangle$, signal by the heterodyne detection $I(t)$ is described by
\begin{equation}
  I(t) =   {\rm Re}E(t)e^{i \Omega_1 t} \sim  -\sqrt{\Omega_i} \left [ \alpha_2 \sin \omega t + \mu_i {\rm Im } \left ( e^{i\omega t} \sum_j \int_0^t \langle \sigma_x^j(t') \rangle e^{i \Omega_i t'}dt' \right ) \right ] ,
  \label{response}
\end{equation}
where the frequency of the reference light is $\Omega_1$.
Since the second term of the rhs of Eq.\ (\ref{response}) is much smaller than the first term, we evaluate the dynamical behavior of the second term after eliminating the basic oscillatory component of $I(t)$.
As a result, the signal on the Stokes light is described by
\begin{equation}
  \Delta I(t) \sim \mu_2\sqrt{\Omega_2}  {\rm Im}  e^{i\omega t} \sum_{j=1}^2 \int_0^t e^{i\Omega_2t'}(\langle \sigma_x^j(t') \rangle - \bar{s}_x(t')) dt',
  \label{signal}
\end{equation}
where $\bar{s}_x(t)$ denotes the expectation value of $\sigma_x(t)$ for $M=1$.

The dynamics of $\sigma_x^j(t)$ is also studied by its equation of motion given by
\begin{equation}
  i{d \sigma_x^j(t)  \over dt} = -i(\nu (a_j^\dagger(t) + a_j(t))+\varepsilon)\sigma_y^j(t),
\end{equation}
where we obtain a set of operators $\sigma_Y^j(t)$ and $a_j(t)\sigma_y^j(t)$ with unknown time-dependence.
Their equations of motion provide us with another set of operators constructed by $a_j$, $\sigma_\alpha^j$, $c_i$, etc. that corresponds to composite operators dressed with various types of fluctuations of electrons/phonons/photons\cite{comp1,comp2}.
Since the equation of motion does not close in any order, and we obtain a series of composite operators by repeating this procedure.
As a result, $\sigma_x^j(t)$ is expanded by a power series of the parameters in the Hamiltonian.
We write down some of the equations of motion order by order, where the typical composite operators connected with the electron-electron and phonon-phonon entanglement are derived.

It helps us understand the role of photons to describe the case with a single electron-phonon system ($M=1$), where the time-dependence of the operators is omitted in the rest of the paper for simplicity.
\begin{enumerate}
\item{The zeroth order equation:}
  \begin{equation}
    i{d c_j \over dt} = [c_j,{\cal H}_1] = \Omega_j c_j +\mu_j \sigma_x.
  \end{equation}
\item{The first order equation:}
  \begin{equation}
    i{d \sigma_x  \over dt} = -i(\nu (a^\dagger + a)+\varepsilon)\sigma_y.
  \end{equation}
\item{The second order equations:}
  \begin{eqnarray}
  i{d(a \sigma_y)   \over dt}  & = & \omega a \sigma_y + {\nu \over 2}(\sigma_y + i\sigma_x) + i a(\nu (a^\dagger + a ) + \varepsilon )\sigma_x \nonumber \\
  & - & 2i(\sum_j \mu_j(c^\dagger_j + c_j) + \lambda )a \sigma_z \\
  i{d \sigma_y  \over dt} & = & i (\nu (a^\dagger + a)+\varepsilon )\sigma_x -2i \left \{ \sum_j\mu_j(c^\dagger_j + c_j)+\lambda \right \} \sigma_z 
  \end{eqnarray}
\item{The third order equations:}
  \begin{eqnarray}
  i{d (a \sigma_z)  \over dt} & = & \omega a \sigma_z + {\nu \over 2} (\sigma_z+1)  + 2i\left \{ \sum_j \mu_j(c^\dagger_j + c_j) + \lambda \right \} a \sigma_y \\
  i{d (a \sigma_x ) \over dt} & = & \omega a \sigma_x +  {\nu \over 2}(\sigma_x + i \sigma_y)  - ia(\nu(a^\dagger + a)+\varepsilon)\sigma_y \\
    i{d (c_j \sigma_z) \over dt} & = & \Omega_j c_j \sigma_z -i \mu_j \sigma_y + 2i c_j \left \{ \sum_k (c_k^\dagger + c_k )+\lambda \right \} \sigma_y \\
  i{d (c_j a \sigma_z)  \over dt} & = & (\Omega_j + \omega) c_j a \sigma_z + {\nu \over 2}c_j (\sigma_z+1)  \nonumber \\
  & + & 2i c_j \sum_k \mu_k(c^\dagger_k + c_k)a \sigma_y + i (2\lambda-\mu_j)c_ja \sigma_y
  \end{eqnarray}
\end{enumerate}
\ \\
Examples of the composite mode operators up to this order are: $\sigma_x$, $a \sigma_y$, $\sigma_y$, $a \sigma_x$, $a \sigma_z$, $a^2 \sigma_x$, $a^\dagger a \sigma_x$, $c_j a \sigma_z$, and $c_j a \sigma_y$.

For two electron-phonon systems ($M=2$), the equations of motion are given by:
\begin{enumerate}
  \item{The zeroth order equation:}
    \begin{equation}
      i{d c_j \over dt} = [c_j,{\cal H}_2] = \Omega_j c_j +\mu_j \sum_{i=1}^2 \sigma_x^i.
      \label{em1}
    \end{equation}
  \item{The first order equation:}
    \begin{equation}
      i{d \sigma_x^i  \over dt} = -i(\nu (a_i^\dagger + a_i)+\varepsilon)\sigma_y^i.
    \end{equation}
  \item{The second order equations:}
    \begin{eqnarray}
      i{d (a_i \sigma_y^i)  \over dt} & = & \omega a_i \sigma_y^i +  {\nu \over 2}(\sigma_y^i + i\sigma_x^i) + i a_i (\nu (a_i^\dagger + a_i) + \varepsilon) \sigma_x^i \nonumber \\
      & - & 2i \left \{ \sum_j \mu_j(c^\dagger_j + c_j) + \lambda \right \} a_i\sigma_z^i, \\
      i{d \sigma_y^i  \over dt} & = & i (\nu (a_i^\dagger + a_i)+\varepsilon )\sigma_x^i -2i \left \{ \sum_j\mu_j(c^\dagger_j + c_j)+\lambda \right \} \sigma_z^i
    \end{eqnarray}
  \item{The third order equations:}
    \begin{eqnarray}
    i{d (a_i \sigma_z^i)  \over dt} & = & \omega a_i \sigma_z^i + {\nu \over 2}(\sigma_z^i+1)  + 2i\left \{ \sum_j \mu_j(c^\dagger_j + c_j) + \lambda \right \} a_i \sigma_y^i \\
  i{d (a_i \sigma_x^i ) \over dt} & = & \omega a_i \sigma_x^i + {\nu \over 2}(\sigma_x^i + i\sigma_y^i)-ia_i(\nu(a_i^\dagger + a_i)+\varepsilon)\sigma_y \\
  i{d (c_j \sigma_z^i) \over dt} & = & \Omega_j c_j \sigma_z^i -i\mu_j \sigma_y^i + 2i c_j \left \{ \sum_k (c_k^\dagger + c_k )+\lambda \right \} \sigma_y^i + \mu_j \sigma_z^i \sigma_x^{3-i} \\
  i{d (c_j a_i \sigma_z^i)  \over dt} & = & (\Omega_j + \omega) c_j a_i \sigma_z^i + {\nu \over 2}c_j (\sigma_z^i+1)  \nonumber \\
  & + & 2ic_j\sum_k \mu_k(c^\dagger_k + c_k)a_i \sigma_y^i + i (2\lambda-\mu_j)c_ja_i \sigma_y^i + \mu_j a_i \sigma_z^i \sigma_x^{3-i},
    \end{eqnarray}
  \item{The fourth order equations:}
    \begin{eqnarray}
      i{d (c_j a_i \sigma_y^i)  \over dt} & = & (\Omega_j + \omega) c_j a_i \sigma_y^i + {\nu \over 2}c_j (\sigma_y^i+i\sigma_x^i) +i c_j a_i  (\nu (a_i^\dagger + a_i)+\varepsilon )\sigma_x^i  \nonumber \\
      & - &   2i c_j (\sum_k \mu_k (c_k^\dagger + c_k) + \lambda ) a_i  \sigma_z^i + \mu_j  a_i   ( i\sigma_z^i + \sigma_y^i \sigma_x^{3-i}) \\
  i{d (a_i  \sigma_z^i \sigma_x^{3-i}) \over dt} & = & \omega a_i \sigma_z^i \sigma_x^{3-i}+ {\nu \over 2} (\sigma_z^i + 1)\sigma_x^{3-i} + 2i \left \{ \sum_k  \mu_k(c^\dagger_k + c_k) + \lambda \right \} a_i \sigma_y^i \sigma_x^{3-i}   \nonumber \\
  & - &  i a_i (\nu (a_{3-i}^\dagger+ a_{3-i}) +\varepsilon )\sigma_z^i \sigma_y^{3-i}
  \label{em2-1}
    \end{eqnarray}
  \item{The fifth order equation:}
    \begin{eqnarray}
  i{d (a_i  \sigma_y^i \sigma_x^{3-i}) \over dt} & = & \omega a_i \sigma_y^i \sigma_x^{3-i} + {\nu \over 2} (\sigma_y^i + i\sigma_x^i)\sigma_x^{3-i} + ia_i (\nu(a_i^\dagger + a_i ) + \varepsilon) \sigma_x^i \sigma_x^{3-i} \nonumber \\
  & - & 2i \left \{ \sum_j \mu_j (c_j^\dagger + c_j) + \lambda \right \} a_i \sigma_z^i \sigma_x^{3-i} - i a_i (\nu (a_{3-i}^\dagger + a_{3-i})+\varepsilon) \sigma_y^i \sigma_y^{3-i}.\nonumber \\
    \label{em2}
\end{eqnarray}
\end{enumerate}
\ \\
Examples of the composite mode operators up to the fifth order are: $\sigma_x^i$, $a_i \sigma_y^i$, $\sigma_y^i$, $a_i \sigma_x^i$, $a_i \sigma_z^i$, $(a_i)^2 \sigma_x^i$, $a_i^\dagger a_i \sigma_x^i$, $c_j a_i \sigma_z^i$, $c_j a_i \sigma_y^i$, $\sigma_x^1 \sigma_z^2$, $a_1 a_2 \sigma_z^1 \sigma_y^2$, and $a_1 a_2 \sigma_y^1 \sigma_y^2$.

The composite modes which induce interparticle correlation are given by $\sigma_x^1 \sigma_z^2$, $a_1 a_2 \sigma_z^1 \sigma_y^2$ and $a_1 a_2 \sigma_y^1 \sigma_y^2$ at the lowest order of the coupling constants.
The first one corresponds to the electron-electron correlation, while the latter two are relevant to the phonon-phonon correlation that appear in the equations of motion for the first time, i.e., at the lowest order.

Equations (\ref{em2-1}) and (\ref{em2}) show that additional composite modes for phonon entanglement, e.g., $a_1 a_2 \sigma_y^1 \sigma_y^2$, are obtained in the presence of the nonadiabatic interaction $\lambda \sigma_x^i$.
We mention that the nonadiabatic interaction also affects the intersystem entanglement in this way, for example.

Though quantitative discussion on the role of each composite mode is left for further studies, we point out that the quantization of light is essential to understand the entanglement generation between remote systems. 
To be more precise, the above-mentioned composite modes are obtained as a result of the commutation relation $[c_j, c_j^\dagger]=1$, i.e., none of them appears in the series of the equations of motion if $c_j$ and $c_j^\dagger$ are regarded as c-numbers.

\section{Calculated Results}

\begin{figure}
 \scalebox{0.55}{\includegraphics*{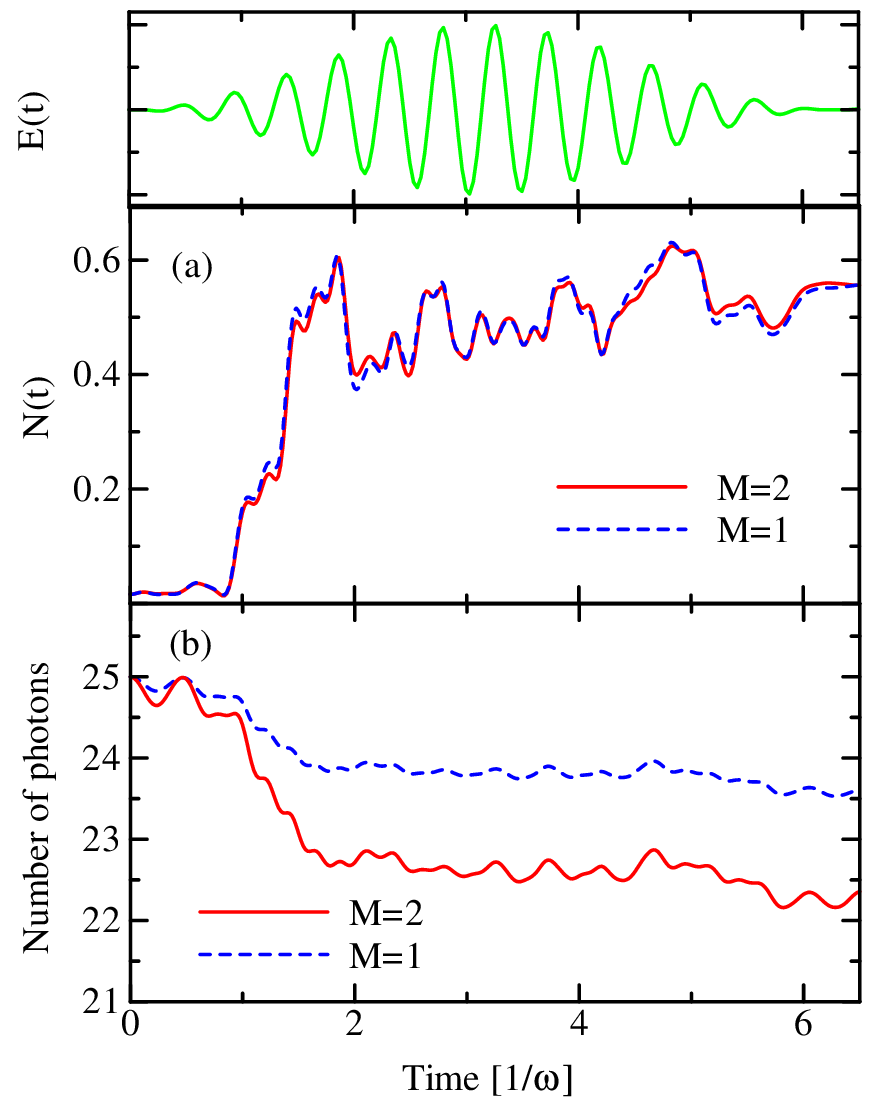}}
\caption{(Color online) Top panel: Electric field amplitude of the incident light pulse. (a) Excited-state population $N(t)$ as a function of time. The red line and the blue dashed line correspond to $N(t)$ for $M=2$ and $M=1$, respectively. (b) The number photons in the pump mode $\langle c_1^\dagger c_1\rangle$ for $M=2$ (red line) and $M=1$ (blue dashed line).}
\label{pop}
\end{figure}

An overview of the dynamics of the system is obtained by monitoring the behavior of observables.
We show the excited-state population $N(t)=\langle \hat{n}_1 \rangle \equiv \langle \Phi(t)|\hat{n}_1|\Phi(t) \rangle$  as a function of time in Fig.\ \ref{pop}-(a), for example.
The blue dashed line in the figure is $N(t)$ for $M=1$, and the top panel of the figure shows the expectation value of the electric field amplitude of the incident pulse $E(t)$.
The two lines in Fig.\ \ref{pop}-(a) almost coincide, as systems 1 and 2 do not interact directly.
The slight difference between them originates from two factors.
One is the difference in the photon states.
As shown in Fig.\ \ref{pop}-(b), the number of absorbed photons linearly depends on $M$, and the photon states also depend on the value of $M$.
Hence, the material systems obey different dynamics as $M$ is varied, which causes the difference of $N(t)$.
We note that this effect becomes negligible as the number of photons increases.
The other is the quantum correlation between systems 1 and 2, which is mediated by irradiated photons.
Obviously, this effect does not exist for $M=1$, which we focus on in this paper.

It is necessary to choose a measure for entanglement in order to discuss the quantum correlation itself.
As we mentioned before, the entanglement between phonons should be referred to as multipartite or mixed-state entanglement\cite{mixed} in the present case.
Among various measures for multipartite entanglement, we calculated the quantum mutual information $I_M(t)$, defined by $I_M(t) = S_A(t)+S_B(t)-S_{A\otimes B}(t)$.
$S_\alpha$ denotes the von Neumann entropy of the reduced density matrix $\rho_\alpha$ on the subsystem $\alpha$\cite{qmi1}.
\begin{figure}
\scalebox{0.55}{\includegraphics*{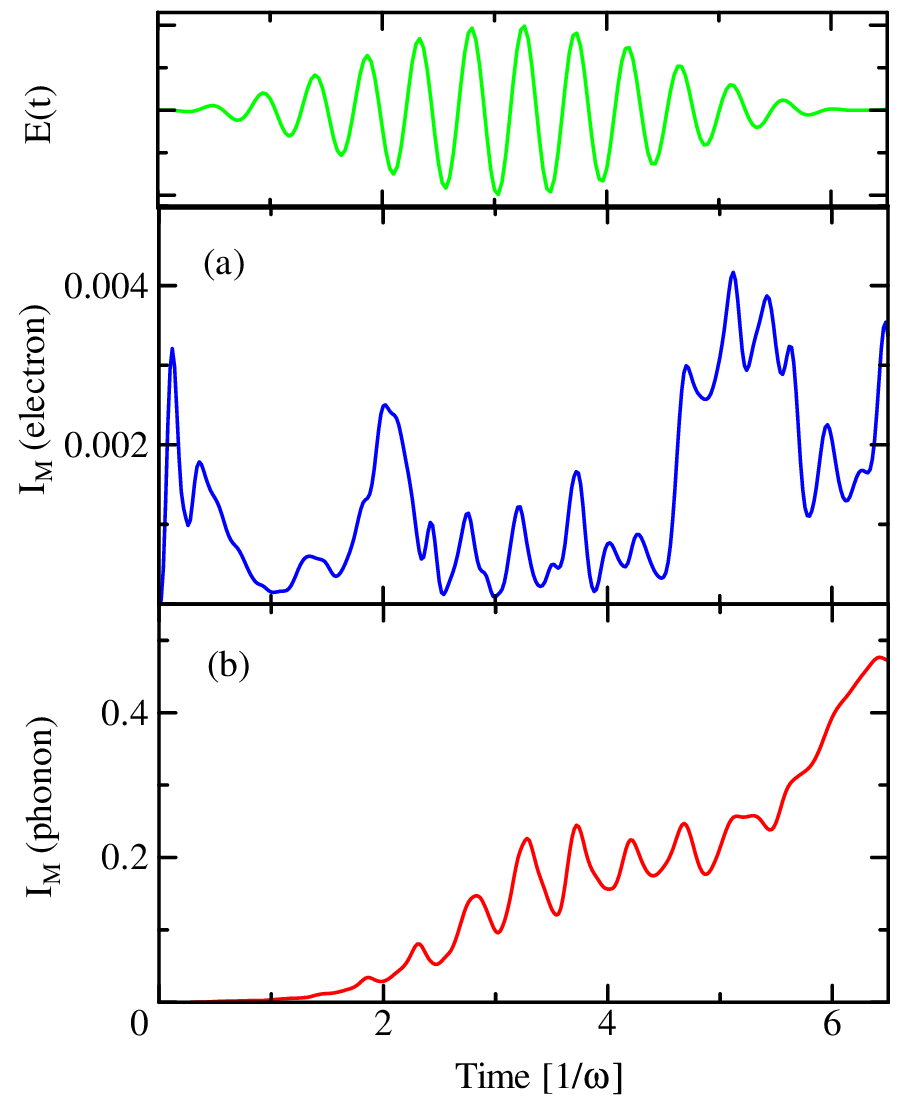}}
\caption{(Color online) Top panel: Electric field amplitude of the incident light pulse. Quantum mutual information $I_M(t)$ for (a) $A, B =$ electrons, and (b) $A,B=$phonons.}
\label{im1}
\end{figure}

Figures \ref{im1}-(a) and (b) show $I_M(t)$ as functions of $t$ for electron entanglement ($A,B=$ electrons) and phonon entanglement ($A,B=$ phonons), respectively.
First, we found that $I_M(t)$ has finite values under pulse irradiation in both cases.
Since the dimension of the phonon Hilbert space is larger than that of the electron Hilbert space in the present study, $I_M$ takes larger value for phonons than for electrons.
On the other hand, regarding the dynamical properties, $I_M(t)$ for electrons increases immediately after the photoirradiation starts, while $I_M(t)$ for phonon entanglement remains at a small value for $t<1$ and starts to increase afterwards.
These features reflect the different entanglement generation mechanisms between electrons and phonons, which we study by considering the composite modes related to the corresponding quantum correlations.

Hence, the time dependence of $\sigma_x^j$ is determined through the motion of these composite modes, and we focus on those which contribute to the quantum correlation between systems 1 and 2 to study the dynamics of $I_M(t)$.
As Eqs.\ (\ref{em1})-(\ref{em2}) show, the operator that contributes to the intersystem quantum correlation first appears in the form of $\sigma_x^1 \sigma_z^2$ in the fourth-order equations of motion.
For phonon entanglement, the corresponding operator is obtained as, for example,  $\sigma_y^1 \sigma_z^2 a_1 a_2$.
These operators do not appear in the case of a single electron-phonon system $(M=1)$, while the other intrasystem operators do.
We also stress that they are obtained only when the quantization of light is taken into account, which reflects the fact that the interaction with the classical electromagnetic field is responsible for the local operations with classical communication (LOCC)\cite{nielsen} in the present configuration.

\begin{figure}
  \scalebox{0.55}{\includegraphics*{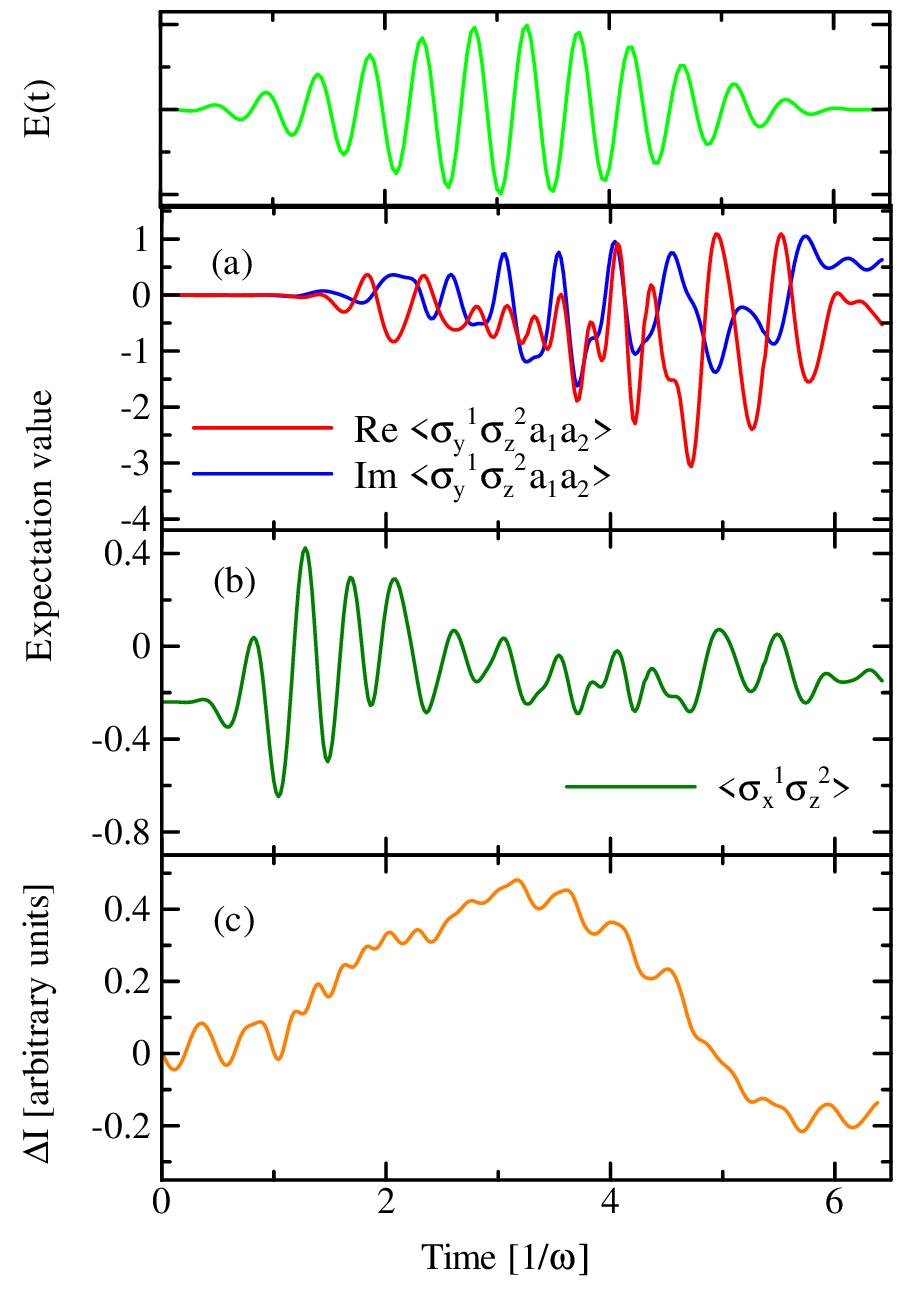}}
  \caption{(Color online) Top panel: Electric field amplitude of the incident light pulse.  Expectation values of the composite operators for (a) $\sigma_x^1 \sigma_z^2$ and  (b) $\sigma_y^1 \sigma_z^2 a_1 a_2$.
    (c) Signal of heterodyne detection $\Delta I(t)$.
    $\langle \sigma_y^1 \sigma_z^2 a_1 a_2 \rangle$ is relevant to the time evolution of $\Delta I(t)$.}
\label{opresponse}
\end{figure}

Figures \ref{opresponse}-(a)-(c) respectively show $\langle \sigma_x^1 \sigma_z^2 \rangle$, $\langle \sigma_y^1 \sigma_z^2 a_1 a_2 \rangle$, and $\Delta I(t)$ as functions of time.
When the incident light is turned on, the time evolution of the system causes the values of these quantities to change.
Regarding the expectation values of the composite operators, $\langle \sigma_x^1 \sigma_z^2 \rangle$ has a finite value even for $t \sim 0$, which contributes to the generation of quantum correlation between electronic states of systems 1 and 2.
Accordingly, $I_M(t)$ for electron entanglement rapidly grows as shown in Fig.\ \ref{im1}.
In contrast, $\langle \sigma_y^1 \sigma_z^2 a_1 a_2 \rangle$, which contributes to the phonon entanglement, grows with a certain delay ($\sim 1$) since it is proportional to the number of excited phonons.

As $c_2$ is expanded by a power series of model parameters, the coefficients of each term are determined by the dynamics of the composite modes.
In this respect, $\langle \sigma_x^1 \sigma_z^2 \rangle$ and $\langle \sigma_y^1 \sigma_z^2 a_1 a_2 \rangle$ belong to the terms of different orders, and thus the dynamical behavior of $c_2$ is not simply determined by their magnitude.
Instead, an important feature shown in Figs.\ \ref{im1}-(b) and \ref{opresponse}-(b) is that the overall behavior of $\langle \sigma_y^1 \sigma_z^2 a_1 a_2 \rangle$ synchronizes with the growth of $I_M(t)$ at the beginning of photoirradiation, which shows that the growth dynamics of $I_M(t)$ for $t < 2$ can be understood by the motion of the relevant composite modes exemplified by $\langle \sigma_y^1 \sigma_z^2 a_1 a_2 \rangle$.
We conclude that the phonon entanglement generation process is composed of two subprocesses corresponding to the dynamics of the relevant composite modes such as $\langle \sigma_x^1 \sigma_z^2 \rangle$ and $\langle \sigma_y^1 \sigma_z^2 a_1 a_2 \rangle$\
The incubation period in $I_M(t)$ is caused by the transition between them, and  we should consider the dynamics of this two-step process to design a coherent control method of phonon entanglement.

Although $E(t)$ is small for $t<1$, photons in each mode interact with electrons and there is a slight change of the number of the photons as shown in Fig.\ \ref{pop}-(b).
As a result, electron entanglement appears when the light irradiation starts.
In contrast, as shown in Fig.\ \ref{pop}-(a), electronic population transfer does not take place when $E(t)$ is small, and the creation of phonons is suppressed during this period of time.
Thus, Fig.\ \ref{im1}-(b) shows that phonon entanglement is generated after a certain phonon population has accumulated in the system.

Figure \ref{opresponse}-(c) shows that $\Delta I(t)$ increases after weak oscillation for $t < 1$.
Since this change of behavior is similar to that in $I_M(t)$ and $\langle \sigma_y^1 \sigma_z^2 a_1 a_2\rangle$, we conclude that the measurement of $\Delta I(t)$ will give information on phonon entanglement, i.e., the heterodyne detection of the Stokes light is suitable for observing the dynamical properties of phonon entanglement generation.

\section{Summary}
In this paper, we studied the dynamics of entanglement generation between remote systems by the irradiation of a quantized light pulse.
Employing a model of coupled electron-phonon-photon systems, we found that the quantum mutual information for phonons $I_M(t)$ reveals the dynamics of phonon entanglement generation.
Since phonons are coupled with electrons, phonon entanglement is mediated by electron entanglement generated by photoexcitation.
Hence, the growth of the electron entanglement precedes that of the phonon entanglement, and $I_M(t)$ for phonons grows with a certain incubation period.
This feature can be understood from the dynamics of composite modes derived from the Heisenberg equations of motion.

We took the coherent states of photons as the initial state.
Since the composite modes for the entanglement generation are found in the equations of motion, the other types of photon states, e.g., the Fock states and the squeezed state, etc., also provide the quantum entanglement between remotes systems.
Quantitative study on the effect of the quantum state of the incident light is left for the future study.

The average number of photons at $t=0$ is $\sum_i|\alpha_i|^2=37.5$.
When the incident light is intense, the semiclassical approximation in which light is regarded as classical field is valid.
Though the light intensity in the present study is weaker than real experimental situations, we obtain similar results for one-body operators such as $\sigma_z^j$\cite{epjdme}.
Hence, the present results show that the quantized electromagnetic field should be taken into account in order to study the quantum correlation between material degrees of freedom, even when the semiclassical approximation seems to be valid.
In other words, the transient effects on quantum many-body states are not clarified so far and the investigation on them are left for future studies.

Detection methods of entanglement generation dynamics were also studied and we showed that scattered light, e.g., Stokes light, carries information on phonon entanglement.
We derived relevant composite modes for the complex amplitude of the Stokes light and found that its heterodyne detection will help us distill the information on phonon correlation, which slowly increases after a certain number of phonons are created.
In this way, the present results show us a way to understand the transient behavior of quantum correlation in interacting many-body systems.

In a previous study of phonon entanglement between remote diamond crystals\cite{lee}, the entanglement is generated as a result of measuring on the scattered light, i.e.,  measurement corresponds to a disentanglement process between photons and phonons.
Under the projection hypothesis, this means that the quantum correlation before measurement is also important for finding appropriate methods for entanglement control.
In other words, the study of the transient behavior of entanglement generation reveals the projected subspaces required to realize designed entangled states.
This point is also important when we consider the coherent control of quantum many-body states.
Since $I_M$ acts as an order parameter of quantum phase transitions\cite{qpt1}, dynamics of $I_M$ shown in the present study will be a basis for further study on the coherent dynamics of photoinduced quantum phase transitions, particularly in the earliest stage of the nucleation\cite{piptme,piptme2}.
The present results show that quantum nature of light-matter interaction plays an important role there, and that the precise generation mechanism needs to be revealed in order to realize the optical control of quantum cooperative phenomena.

\acknowledgments
K.I. is grateful to K. G. Nakamura for fruitful discussion.
This work was partly supported by the research funding granted by Utsunomiya University President, JSPS KAKENHI Grant Numbers JP18K03456 and JP18K03474, and the Collaborative Research Project of Laboratory for Materials and Structures, Institute of Innovative Research, Tokyo Institute of Technology, Japan.
Numerical calculations were performed on the facilities of the Supercomputer Center, Institute for Solid State Physics, the University of Tokyo, Japan, and the computer resources offered under the category of General Projects by Research Institute for Information Technology, Kyushu University, Japan.

\end{document}